\documentclass[sigconf]{acmart}

\usepackage{booktabs} 
\usepackage{todonotes}
\usepackage{csquotes} 
\usepackage{enumitem} 

\newcommand{\code}[1]{\mbox{\texttt{#1}}}
\newcommand{\shrink}{\vspace{-1ex}}
\newcommand{\sshrink}{\vspace{-1.5ex}}

\setcopyright{rightsretained}

\acmConference[SEMANTiCS'17]{13th International Conference on Semantic Systems}{September 2017}{Amsterdam, Netherlands} 
\acmYear{2017}
\copyrightyear{2017}

\acmPrice{1/3}

\begin{document}
\title[Good Applications for Crummy Entity Linkers]{Good Applications for Crummy Entity Linkers?
The Case of Corpus Selection in Digital Humanities}

\author{Alex Olieman}
\orcid{0000-0001-6533-5328}
\affiliation{%
  \institution{University of Amsterdam}
}
\email{olieman@uva.nl}
\affiliation{%
  \institution{Stamkracht BV}
}
\email{alex@stamkracht.com}

\author{Kaspar Beelen}
\affiliation{%
  \institution{University of Amsterdam}
}
\email{k.beelen@uva.nl}

\author{Milan van Lange}
\affiliation{
  \institution{NIOD Institute for War, Holocaust, and Genocide Studies}
}
\email{m.van.lange@niod.knaw.nl}

\author{Jaap Kamps}
\orcid{0000-0002-6614-0087}
\affiliation{%
  \institution{University of Amsterdam}
}
\email{kamps@uva.nl}

\author{Maarten Marx}
\affiliation{%
  \institution{University of Amsterdam}
}
\email{maartenmarx@uva.nl}

\renewcommand{\shortauthors}{A.\ Olieman et al.}

\begin{abstract}
Over the last decade we have made great progress in entity linking (EL) systems, but performance may vary depending on the context and, arguably, there are even principled limitations preventing a ``perfect'' EL system. 
This also suggests that there may be applications for which current ``imperfect'' EL is already very useful, and makes finding the ``right'' application as important as building the ``right'' EL system.
We investigate the Digital Humanities use case, where scholars spend a considerable amount of time selecting relevant source texts.
We developed \textit{WideNet}; a semantically-enhanced search tool which leverages the strengths of (imperfect) EL without getting in the way of its expert users. We evaluate this tool in two historical case-studies aiming to collect a set of references to historical periods in parliamentary debates from the last two decades; the first targeted the Dutch Golden Age, and the second World War II. 
The case-studies conclude with a critical reflection on the utility of WideNet for this kind of research, after which we outline how such a real-world application can help to improve EL technology in general.

\end{abstract}

%
%

\maketitle

\shrink
\section{Introduction}
The traditional approach for evaluating Entity Linking (EL) systems employs metrics borrowed from Information Retrieval, \textit{precision} and \textit{recall} \cite{Erp2016,Wang2015,Hasibi2015}, to summarize system performance and to aid in their comparison. Consequently, the notion of improving EL technology is closely tied to demonstrating an increased performance in terms of precision and recall. Recent evaluation efforts, utilizing multiple benchmark datasets, have however shown that a good performance on one dataset often does not generalize to others \cite{Ling2015,Jha2017}, even if the input texts originate from closely related genres and domains \cite{Hasibi2015,Derczynski2015,Ilievski2016}.

This limitation of current EL benchmarks, we argue, limits the usefulness of performance statistics for application developers who would like to include EL in their software. From their perspective, the current evaluation paradigm is unsatisfactory because it does not provide clear expectations about how the EL functionality will be perceived by their intended users. The current situation is reminiscent of the discussion of Machine Translation evaluation in the early 1990s, which recognized the importance of identifying good niche applications for state-of-the-art technology, which was arguably crummy at the time \cite{Church1993}.
                                                                       
In this paper, we briefly review current-day EL evaluation and discuss the parallels with early work on Machine Translation. We take the position that in order to demonstrate the value of EL for real-world users, a complementary evaluation paradigm based on end-user applications is needed. One particularly high-payoff niche application is found in the area of Digital Humanities, where scholars spend a considerable amount of time selecting relevant source texts, which together form the object of their research.

To assist scholars and other researchers with corpus selection, we have developed \textit{WideNet}; a semantically-enhanced search tool which leverages the strengths of (imperfect) EL without getting in the way of its expert users. As a means of evaluating this tool, two historical case-studies have been conducted. Both aimed to collect a set of references to historical periods in parliamentary debates from the last two decades; the first targeted the Dutch Golden Age, and the second World War II. The case-studies conclude with a critical reflection on the utility of WideNet for this kind of research, after which we outline how such a real-world application can help to improve EL technology in general.

\shrink
\section{Entity Linking Evaluation}
The cornerstone of EL evaluation, as it is reported in the literature, is the use of annotation-based performance metrics, such as \textit{precision} and \textit{recall}. Analogous to their original definitions in Information Retrieval, precision indicates the fraction of generated links that are correct, and recall indicates the fraction of all gold-standard entity links that have been generated by the system \cite{Wang2015,Hasibi2015}. These metrics appeal to the desire for a straightforward evaluation of EL systems, and are highly successful in that sense.

Several issues which arise when these basic evaluation metrics are operationalized in practice have, however, been identified in recent years. In this section we review outstanding issues with EL evaluation, and proceed into our call for a complementary evaluation paradigm which addresses the utility of EL technology for end-users.

\sshrink
\subsection{Annotation-based Metrics}
The common-sense argument in favor of annotation-based evaluation metrics goes as follows: ``If our aim is to generate correct links from mentioned entities in text to their corresponding entries in a knowledge base, one should evaluate an entity linking system on the basis of the annotations it produces.'' Three objects are needed to perform such an evaluation: 1. an entity linking system, 2. a target knowledge base (KB) to link into, and 3. a benchmark dataset, comprised of input texts and corresponding gold-standard annotations. Comparative evaluation campaigns can be organized by specifying which KB and benchmark to use, in order to let multiple EL systems compete against each other with a certain degree of fairness.

Decisions regarding the target KB and the creation of gold-standard annotations are, in practice, made by benchmark creators, which, together with the specific metric that is used, form an operational definition of an entity linking task. For instance, should links be specified at the level of a document, or at the level of an individual phrase? Will the text offsets or boundaries of the links (indicating the entity mentions or surface forms) be provided to the systems or do they have to be identified before disambiguation? Should annotations be generated for entities that are not in the KB, or should these mentions not be linked at all? Different answers to these questions have led to the definition of several incompatible metrics \cite{Ling2015}.

To indicate what the score of a system on a particular benchmark actually signifies, detailed annotation guidelines are required \cite{Erp2016,Ling2015,Jha2017}. Such guidelines should specify, for instance, whether exclusively named entities should be linked, or if common entities and concepts are to be annotated as well, and to what extent \cite{Erp2016,Ling2015}. The guidelines need to be clear about any cases in which multiple annotations on the same mention are acceptable, or even encouraged \cite{Erp2016,Ling2015,Hasibi2015}. However, even when the minute details of `what counts as a good link' are fully specified, it is still exceedingly difficult to draw inferences about EL systems across benchmarks \cite{Ling2015,Erp2016,Jha2017,Wang2015}. By keeping our metrics as simple as possible, and hiding the complexity of the problem definition in guidelines, the burden of understanding the differences between benchmarks falls on those who need to compare between EL systems.

These issues are compounded by differences in genre and domain of the text that EL is applied to. Most benchmark datasets are compiled from a sample of documents drawn from a specific collection \cite{Erp2016}. The characteristics of the source collection are apparent in the distributions of lexical forms and entity mentions, and have a significant influence on the performance of systems that exploit them. Some genres, notably microblog posts and search queries, are so different from the types of documents that EL is commonly applied to that they are considered separate variants of the Entity Linking task \cite{Hasibi2015,Derczynski2015,Cornolti2016}. Similarly, to effectively perform EL in specific domains, many researchers have used specialized KBs and needed to create new benchmarks to meet their goals, e.g. \cite{Olieman2015a,Thorne2016,DeWilde2015}.

Domain-agnostic benchmarks, however, present their own set of issues. Current datasets are skewed toward popular, frequently-mentioned entities, and contain many surface forms that have a low degree of ambiguity within the dataset \cite{Erp2016,Ilievski2017}. Even simple (baseline) EL systems can achieve a high performance on these benchmarks by heavily weighting popularity features. For example, when such a system processed the surface form ``Ronaldo'', it would be unduly rewarded for considering only the top-5 most frequently mentioned Ronaldos at the time \cite{Ilievski2017}. If a high score can be achieved by solving a simplified version of the EL problem on skewed datasets, it should not be mistaken for an indicator of real-world performance.

Much valuable work remains to be done in this area. At the least, EL benchmarks should be clearly documented, and would be more useful if they incorporated text from different domains and publication times, featuring a variety of long-tail entity mentions \cite{Erp2016,Ilievski2017}. Ilievski et al. recommend that
\blockcquote[p. 5]{Ilievski2017}{systems should be evaluated on long-tail entities by either introducing new long-tail dataset(s) that deliberately contain entities with low dominance and high ambiguity; or by focusing on the few ``hard'' cases in current datasets.}

\sshrink
\subsection{Real-world Applications}
There is a poor understanding of the state-of-the-art in Entity Linking \cite{Erp2016,Wang2015,Ling2015,Ilievski2016a}, which can be partly blamed on the desire for generality in the task definition and its associated objects. How might we then go about demonstrating how effective or useful EL technology can be, in spite of the issues with our metrics and benchmarks? It would be beneficial, in terms of impact and continued funding, to show that EL technology can already be valuable for end-users in its current form.

When the applications of Entity Linking are described, it tends to be in terms of what EL can contribute to other tasks in Natural Language Processing. In \cite{Wang2015} an overview is given of demonstrated benefits for Information Extraction, Knowledge Base Population, Information Retrieval, Content Analysis, and Question Answering tasks. There are far fewer published examples that explicitly address the benefits to end-users in their aims and evaluation, such as \cite{Olieman2016,Odijk2013,DeWilde2015}. An important advantage of doing so, is that proof-of-concept applications are more suited than benchmark results to demonstrate \textit{what EL technology can currently do}. Benchmarking evaluations, we argue, are essential in order to challenge our systems by testing on hard cases, to raise the bar for ourselves. Real-world applications, in contrast, show a greater potential for demonstrating to other stakeholders what our current technology is good for.

Church and Hovy made the same kind of argument in the early 1990s about Machine Translation evaluation \cite{Church1993}. They identified parallels with the Speech Recognition community, pointing out that both communities, at some time, suffered from unrealistically high expectations after a good performance was shown in an overly simplistic evaluation setting. The subsequent failure to meet those expectations, in both cases, led to a decrease in funding of work in these areas \cite{Church1993}. The speech community recovered from this episode by managing expectations more appropriately, and by making progress on several niche applications that would turn out to be quite commercially attractive \cite{Church1993}.

\shrink
\subsection{Identifying a Good Niche Application}
There are several advantages to focusing on niche applications that show the technology in a good light. Not all applications are equally suitable, however. Church and Hovy enumerate six desiderata that we may do well to embrace in the EL community:
\renewcommand{\labelenumi}{\alph{enumi}.}
\blockcquote[p. 246]{Church1993}{Of course, we need to find good niche applications that really do have
value and avoid bad ones that only look as if they will scale up to something important, but actually don't. A good niche application \textelp{} should meet as many of the following desiderata as possible:
  \begin{enumerate}
    \item it should set reasonable expectations, 
    \item it should make sense economically, 
    \item it should be attractive to the intended users,
    \item it should exploit the strengths of the machine and not compete with the strengths of the human,
    \item it should be clear to the users what the system can and cannot do, and
    \item it should encourage the field to move forward toward a sensible long-term goal.
  \end{enumerate}
}

\shrink
\section{Searching for Entities by the Dozen}
We have found a niche application within the field of Digital Humanities which, we argue, helps to showcase current-day EL technology by supporting scholars in complex search tasks.

\sshrink
\subsection{Semantically-enhanced Search}
Applying semantically-enhanced search to large digitized text collections in the humanities was first proposed by Hinze et al. \cite{Hinze2015}. It aims to address the gap between the research questions and methods of the humanities and full-text (lexicographic) search; that of conceptually-based information needs on one side, and  term-based inverted indexes on the other. Toward this end, the documents are processed by a semantic annotation system, which links concepts and entities as they are mentioned in the text to a KB, and these links are incorporated into the existing search indexes \cite{Hinze2015}. Secondly, the semantically-enhanced search system should provide an interface that is suitable for humanities scholars, thereby allowing \blockcquote[p. 149]{Hinze2015}{technical non-experts to use semantic technology when querying document corpora that do not provide \textins{the required} semantic mark-up.}

The concepts that historians use to focus their research are particularly challenging search targets. Periods such as the `Dutch Golden Age,' and `World War II' have a complex and heterogeneous character, and are comprised of a myriad of entities of different types \cite{Shaw2013}. To help historians and other scholars find references to such complex search targets in diachronic corpora, we have developed WideNet, a semantically-enhanced search tool which leverages the strengths of current-day EL while leaving important decisions about relevance in the hands of its users.

\sshrink
\subsection{Offline Processing}
We had already generated entity links for a collection of parliamentary proceedings using DBpedia Spotlight, as described in \cite{Olieman2015a}. For WideNet, we haven't tuned the EL system, to emphasize that imperfect entity linkers can provide value to end-users. The DBpedia Spotlight annotations, with an estimated precision of 0.69 and recall of 0.40 \cite{Olieman2015a}, are incorporated into existing search indexes as additional (nested) fields.

As a source of mappings from complex concepts like `Dutch Golden Age' to more specific concepts and entities, we load a subgraph of DBpedia, corresponding to Wikipedia's category network and its related entities, into a property graph database (see \cite{Rodriguez2015}). The category network is used at runtime to select potentially relevant entities given a root category, by traversing \code{dct:subject}\footnote{For namespace prefixes, see \url{https://dbpedia.org/sparql?nsdecl}.} and \code{skos:broader} relations in reverse direction. Our proof-of-concept makes use of DBpedia, but any KB that conforms to the SKOS ontology can in principle be loaded.

Finally, the system needs access to coarse temporal clues about entities. Because DBpedia does not provide this data reliably across entity types, we extract mentioned years from the \code{rdfs:comment} values of DBpedia resources with a simple regular expression, and add them to the graph.

\shrink
\subsection{Search Interface Design}

\begin{figure}
  \includegraphics[width=.45\textwidth]{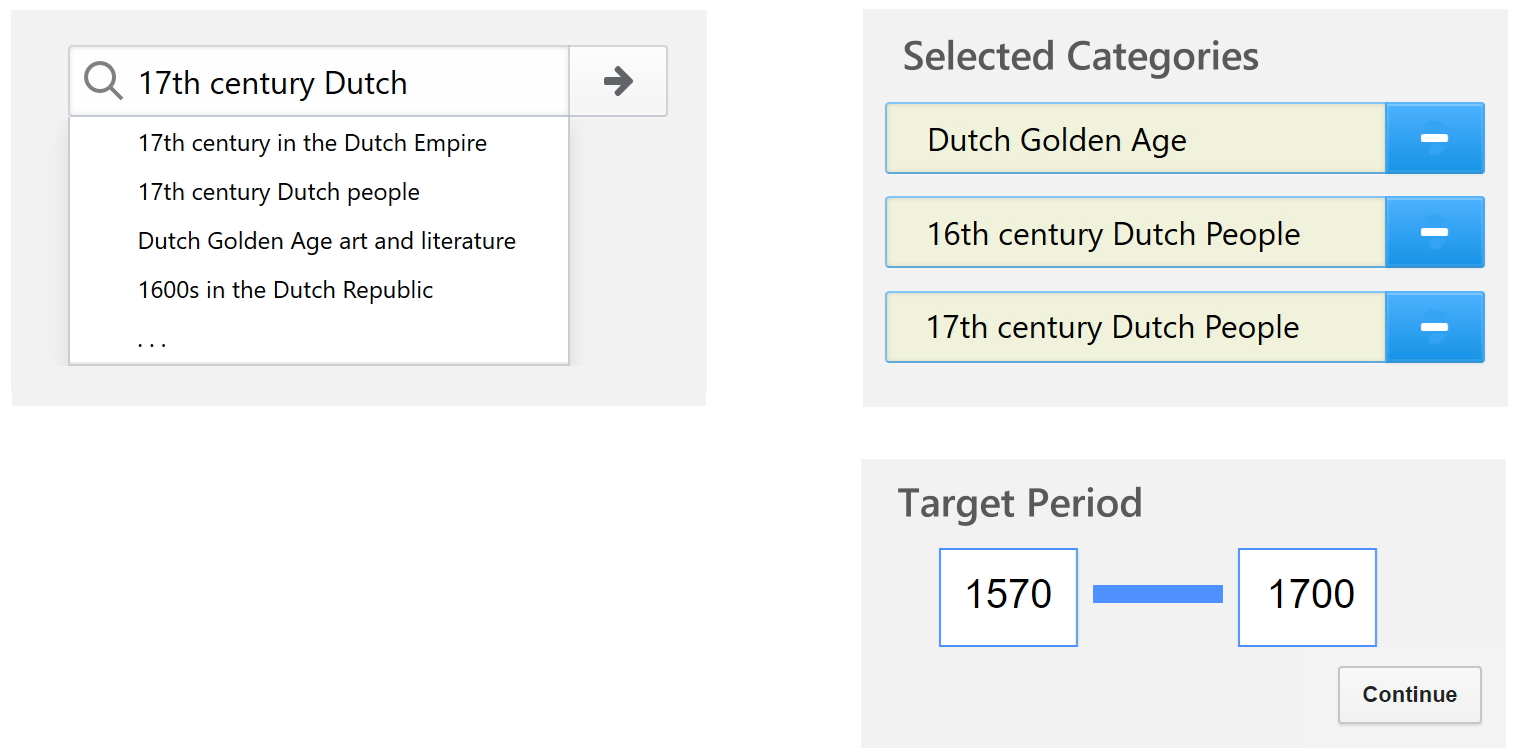}
  \caption{Initial query specification in WideNet.}
  \label{fig:ga-root}
\end{figure}

\begin{figure}
  \includegraphics[width=.45\textwidth]{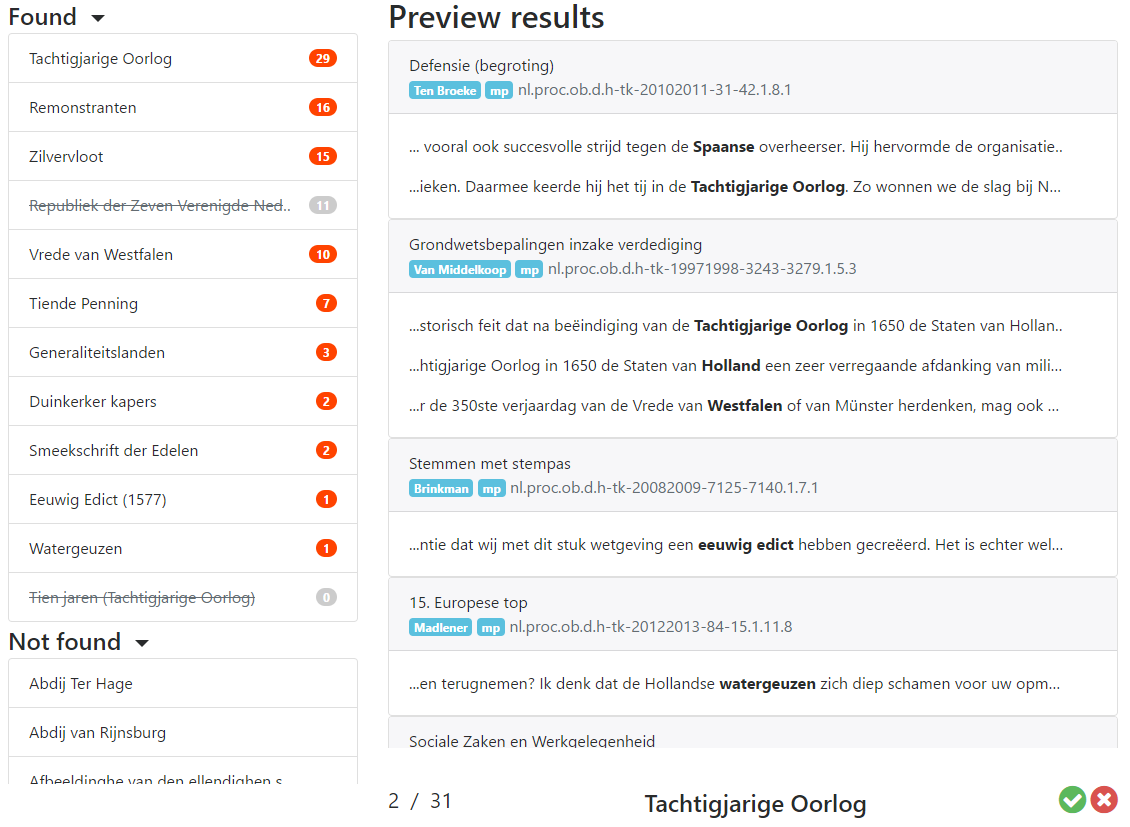}
  \caption{Assessing the relevance of categories and entities.}
  \label{fig:ga-preview}
\end{figure}

\begin{figure}
  \includegraphics[width=.45\textwidth]{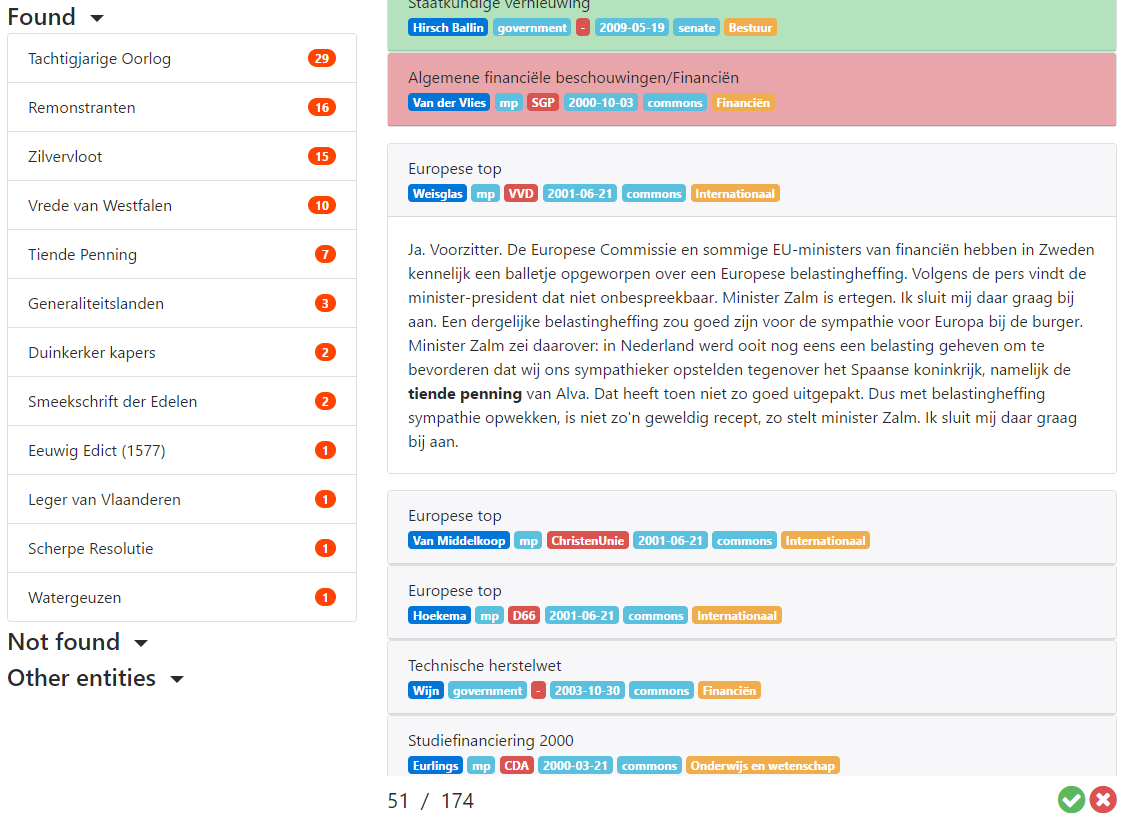}
  \caption{A closer look at the retrieved documents.}
  \label{fig:ga-reading}
\end{figure}

The scholar, using WideNet, starts by selecting one or several root categories from a typeahead search box (see Figure \ref{fig:ga-root}), and can further demarcate the query by selecting a time period, which is used to prune the underlying entities of the selected categories.
WideNet subsequently retrieves the network of narrower categories for each selected root category, and collects the contained entities as potentially relevant query components. Behind the scenes, each entity is compared with the target period, and is considered to be outright relevant to the period, or not, or a borderline case, or as lacking temporal clues altogether. In the current implementation this classification is achieved with simple rules, based on the features: `fraction of years within period,' `fraction of intervals that overlap with the period,' and `has at least one year in period.'  The system uses this information to deselect (sub)categories where more than half of the dated member entities are out-of-period, i.e., those categories are excluded from the query.
%
%

The next step for the WideNet user is to assess which of the retrieved subcategories actually contain entities that lead to relevant results. The interface facilitates this task by showing, per subcategory, which entities are mentioned in the corpus, and how frequently, as well as which entities did not occur (see Figure \ref{fig:ga-preview}). It also displays a list of preview results, showing limited context, to offer quick clues about the relevance of the category. This preview is also useful to identify individual entities that are not relevant after all, which can be deselected by the user.

After inspecting and selecting relevant categories of entities---thereby sculpting the final query, the WideNet interface allows further exploration by providing an environment in which the retrieved documents can be studied up-close, as shown in Figure \ref{fig:ga-reading}.  By situating the close reading activity within the same interface, the user is able to compile a corpus of relevant documents which may be saved and exported.
Moreover, the user can examine the results in relation to the document metadata, e.g. to look for saliency by plotting the annotations over time, or study bias by comparing how often different political parties refer to the entities of interest.

\shrink
\section{Corpus Selection Case-studies}
With the advent of the `linguistic turn,' the analysis of discourse became part and parcel of the historian's toolkit.
Even though the rapid digitization of historical texts promised to bring about a paradigm shift in the humanities, the research methods and heuristics of historians actually did not budge: they generally preferred a qualitative study on a small but carefully curated corpus, while relying on traditional, manually composed indices for finding relevant documents. 
The access to machine-readable text, however, opens up new opportunities for navigating and analyzing the textual traces of the past. 
WideNet purports to assist historians with the digital exploration of vast corpora, enabling them to ``holistically assess the typicality, scope, and power of key issues'' \cite{Blaxill2013} across huge text collections. 
It thereby leverages full-text access to introduce a more data-driven approach to corpus selection in the humanities. 
In what follows, we demonstrate how WideNet actually assists scholars with longitudinal research by elaborating on two ongoing research projects which are based on the verbatim records of the debates in the Houses of Parliament. 

\shrink
\subsection{Background}
Both case studies draw on the digitized and enriched version of the Dutch parliamentary proceedings. 
As part of the PoliticalMashup (and later Dilipad) project, these proceedings were provided with structural semantic annotations, identifying, for example, the party affiliation of the member of parliament (MP) taking the floor, and the title of the debate a speech was part of.
%
%
Moreover, the data was linked to various external (Linked Open Data) resources and knowledge bases, such as DBpedia.
This endeavor greatly stimulated historical research on Dutch (parliamentary) politics. 
The proceedings contain debates that touch almost on every issue that moved Dutch public opinion over more than two centuries.
Despite its centrality in the political landscape, the proceedings' unwieldy size made it difficult for historians to explore---all the more because the existing indices were rather limited in scope.
Digitization and enrichment have unlocked this resource in novel ways. 
%
%

\shrink
\subsection{Dutch Golden Age}
We applied WideNet to study the changing face of Dutch nationalism from the late 1990s to the present. We gauged the role of heritage in reviving contemporary national identities by analyzing historical references to the so-called ``Golden Age.'' The question originated from recent debates on in the changing discourse about ``Dutchness.''
Since the end of the twentieth century, observers (inside as well as outside academia) argued that the Netherlands experienced a rapid transition from a ``thin'' to a ``thick'' conception of national identity. The rather abrupt return to nationalism saw a more procedural and heterogeneous perception of Dutchness being substituted by a culturalist, homogenizing version \cite{tonkens2008culturalization}. 

The thickening of national identity became apparent in various policies and narratives that pitched ``We'' against ``Them.''  Skepticism toward the EU,  stricter immigration laws, and rejection of multiculturalism all fit within this trend.
Also, the establishment of a historical canon---a unified narrative for teaching history in primary and secondary schools---shows how molding the past supported a culturalist understanding of national identity.

\shrink
\subsubsection{Motivation}
In this case study, we assessed the weight of the past on the perceived resurgence of nationalism, by scrutinizing the references politicians in parliament made to one of the most celebrated eras in Dutch history, the so-called ``Golden Age.'' Since the 19th century, this era has served as a benchmark (``ijkpunt'') of national identity \cite{jensen2012gouden}. As the apex of Dutch civilization, the Golden Age has played a crucial role in defining the Dutch national ``We.'' Events and individuals from this era came to symbolize wider national characteristics: the bravery of the ``Geuzen,'' the enlightened thinking of Spinoza; such historical ``memes'' together compose a rich resource for identity construction \cite{de1999discursive}.
%
%

We set out to scrutinize the role of the Golden Age in contemporary Dutch political discourse\footnote{To try this query with WideNet: \url{http://widenet.politicalmashup.nl/nl/preview/ge/}}, as means to empirically assess how history played out in the recent transformations of Dutch politics.
Towards this end, we inductively established the overarching argumentation patterns in which references to the Golden Age appeared. We distinguished different spheres of identity construction: the economical, cultural and political realm---which flowed logically from Wikipedia's Golden Age categorization, including entities predominantly related to trade (Dutch East India Company or VOC), political events (Eighty Years' War), and culturally significant persons (such as Spinoza). After grouping these references we study the selected speeches in relation to other metadata such as party and time. Through a close reading, we assessed how these mentions tie in (or not) with processes of identity formation and contestation. Specific linguistic elements such as pronouns and verbs served as analytical instruments to uncover the way political language ``bounds and bonds people'' into distinctive groups. 

\shrink
\subsubsection{Findings}
From a temporal perspective, the number of Golden Age related references did not show a clear trend. We observed a slight increase after the turbulent period between 2002 and 2004, but differences remained small overall. 
Aggregating Golden Age references by party revealed more fundamental discrepancies. Clearly, the discourse of the conservative and populist right showed a greater tendency to invigorate their nationalistic appeals with historical references. The Reformed Party and the Party for Freedom (PVV), were proportionally the most active in this respect, followed by the right-liberal People's Party for Freedom and Democracy (VVD).
This applies even more to the political references, which almost exclusively circulated on the right. Economics, on the other hands, showed a more balanced distribution, skewing even a bit towards the left.

By aggregating the selected speeches by annotations and metadata fields, WideNet enables researchers to conveniently map their data along different dimensions, thus disclosing general patterns in the discourse about the Dutch seventeenth century. The fact that most of these mentions seem to stem from benches at the right of the spectrum, suggests that, indeed, the Golden Age functioned as an instrument for forging a more exclusive, culturalist understanding of Dutchness.

However, a close inspection of the speeches that survived the filtering process (as shown in Figure \ref{fig:ga-reading}), yields a more nuanced reading and sheds light on the actual (and often subtly) linguistic strategies that politicians use to forge a national unity. 
The political entities mainly comprised wars and taxation, such as the Eighty Years' War and the ``tiende penning,'' a VAT on movables, introduced by the Duke of Alva.
Linguistically, the use of personal pronouns, such as ``we'' became apparent. 
Verheijen, an MP of the right-liberal VVD, reacted negatively to proposals which would give Europe more leverage in national taxation matters. To support his argument he asserts that ``we waged 80 years of war against Spain to obtain our independence.'' 
Verheijen insert a transhistorical ``we'' in his speech,
He amalgamates the Dutch who fought Spain with the Dutch today, who resist infringement on their national independence by Europe.
Generally, MPs on the right exhibit a greater intimacy with the past, relying more heavily on the ``we'' as a homogeneous national actor, or using a cognitive verb, suggesting to have direct access to the minds of the illustrious Dutchmen of the past. Another example of this is Madlener of Wilders' PVV, who insists that ``we won against Spain''---a reference to the Eighty Years' War during the seventeenth century. Moreover, he exclaims: ``I think the Sea Beggars are deeply ashamed of your remark'', thereby rejecting the claims of an opponent by straightforwardly probing and exposing the minds of actors from the distant past.

While the memory of political events was dominated by the right,  discussion of the economical aspects of the Golden Age was more evenly distributed, but divided in diametrically opposite narratives.
The interpretation of the Dutch colonial history and its trade practices figures here as the main bone of contention. 
This debate was sparked by a remark of then-prime-minister Jan Peter Balkenende, who urged the Dutch to embrace again the ``VOC-mentality,'' which he characterized as a tradition of risk-taking and brave, global entrepreneurship. 
The depiction of the VOC as the flagship of Dutch capitalism---based on the values Freedom, Entrepreneurship, and Competition, according to Ten Broeke of the VVD---bounced against a wall of skepticism raised by left-wing MPs, who mostly emphasize the exploitative practices. Vendrik of GreenLeft, for example posits that the VOC mentality actually means ``becoming rich at the expense of others'', while Irrgang, a member of the Socialist Party, equates the VOC---and the underlying mentality---with ``nothing more than colonial plundering, the creation of monopolies.''
%
%

\sshrink
\subsection{World War II}
The second case study is part of the NIOD research project `War and Emotions.' This project focuses on emotional expressions related to World War II in historical Dutch texts. 
It investigates the emotional perspectives on the war as a historical event in, for example, post-war Dutch parliamentary debates.
Whereas emotional legacies of World War II are usually investigated with a focus on case studies or with a purely qualitative approach, this project's aim is to investigate affect in texts from a qualitative and computational perspective by combining close reading with a more quantitative approach. 
The availability of a sufficiently large historical dataset, such as the digitized parliamentary proceedings, offers promising opportunities to subject emotions in relation to speaking about World War II to a fundamentally new assessment.

\sshrink
\subsubsection{Motivation}
The first step in this research project was the selection of relevant debates from the Dutch parliamentary proceedings, to serve as a corpus for quantitative text analysis. 
In this case study, we distinguished debates dealing with directly World War II-related subjects (such as the conviction or release of war criminals) from those in which MPs refer to the war without it being the actual focus of the discussion. 
Finding the former seems relatively straightforward: A fair amount of war debates can be gathered by looking for evident war-related terms (e.g. `war,' `resistance,' `war criminal') in the debate titles in the PoliticalMashup dataset. This process can even be performed manually by a well-informed historian, by closely reading the titles of \textasciitilde{}25 thousand debates. 

The latter task, however, proved more difficult, since it required us to collect the often short references that are scattered throughout the proceedings. 
Keyword search was too impractical a tool for harvesting war-related references and comparisons.
Historical episodes as versatile and complex as World War II are not easily captured with a handful of search queries. 
WideNet facilitated this task\footnote{To try this query with WideNet: \url{http://widenet.politicalmashup.nl/nl/preview/wo2/}}. 

As part of the evaluation of WideNet we focus on three problems:
\begin{enumerate}[leftmargin=*]
	\item References to the war can take a wide variety of forms, e.g. `World War II,' `the Occupation,' `1940-1945'
    \item A plethora of aspects or components of the war are involved in war-related references: `Adolf Hitler,' `Holocaust,' and all sorts of evidently war-related events, people, and places.
    \item In both everyday language and in parliamentary language use, `World War II' is often only used as an indication of time (e.g. before, since, after) without saying anything about the historical event itself. Such time indications are irrelevant to us.
\end{enumerate}

\shrink
\subsubsection{Findings}
WideNet proved to be especially valuable and useful in expanding the World War II-references corpus. 
It helped to gather a corpus of speeches on a variety of topics, containing war-related remarks, comparisons, and references, even if the reference consists of only one or two sentences referring to a war-related topic. After close reading the documents found by WideNet, references containing terms that are unambiguously World War II-related, such as `Jappenkamp' or `Nazi,' proved to be most relevant. 

One of the main limitations of the tool, is that a lot of the mentions of war-related persons, battles, concentration camps, and weapons that WideNet initially included into the query, were irrelevant because they occurred in a in non-war-related context. In this case study, we can distinguish three categories of error:

The first example consists of the terms that occur frequently in a war-related context, but are not characteristic or unique for war-related speeches or references, such as the word  `loyaliteitsverklaring' (loyalty statement) or the acronym `WA'---the latter may stand for `Wehrmacht' but is also used to refer to `Wettelijke Aansprakelijkheidsverzekering' (mandatory liability insurance).

The second category is formed by mentions of persons who were politicians, collaborators, or members of the Dutch Resistance. Mentions of their names can not necessarily be marked as a war-related reference. An example is the occurrence of the name of former Resistance member Marge Klomp\'e, who became the first female minister in the Netherlands in the 1950s. Another example is the name of Simon Carmiggelt, who was a former Resistance member who became famous as writer after the war. Moreover, the same applies to names of resistance newspapers that became mainstream after the war, such as `Trouw.'

The third category is formed by parts of word and/or name combinations that are only World War II-related when used in combination with other names or words. An example is the name `Anne Frank.' This name is especially war-related, but consists of a combination of two very common Dutch names. A fair part of the WideNet search results are established by occurrences of just one of these names. Of course, the names `Anne' and `Frank' are used very often in a non-war-related context. The same applies to names of battles or concentration camps, which very often partly consist of a toponym, such as the names of the Dutch concentration camps `Kamp Amersfoort' and `Kamp Vught.' This problem is caused by a common disambiguation error.

\sshrink
\subsection{Reflection}
In what follows we reflect on how useful WideNet is for corpus selection, but also engage with a broader discussion on how semantically-enhanced search technology supports academic research in the humanities. How does this type of search compare to existing interfaces? To what extent do recall and precision offer valuable metrics for academic search and how should their trade-off be perceived?  We structure this discussion by following the desiderata of Church and Hovy for ``good niche applications'' \cite{Church1993}.

\shrink
\subsubsection{It should set reasonable expectations}
The ``Golden Age'' as well as ``World War II'' function as container concepts for a wide range of spatiotemporally anchored entities. Searching for references to these entities in the parliamentary proceedings amounts to a comprehensive needles-in-haystack problem since the target of our search is scattered for two reasons. Firstly: There is no debate ``about'' the Golden Age, as the topic merely survives by virtue of association or analogy, i.e. mentions conjured up during debates on often totally unrelated topics.
Secondly: spanning more than one hundred years, the Golden Age comprises a wide range of events and personae. Even though the Second World War covers ``just'' six years, our knowledge about this period vastly exceeds the Dutch seventeenth century, and an immense amount of entities are related to this event as well.
Faced with such complexity, expectations about the performance should be evaluated against relevant baselines, the existing tools as well as manual approaches.

For sure, semantic annotation technology does not have the accuracy of manual curation.  
In terms of quality, computational approaches are not (yet) competitive with meticulously curated indices. But the time required for manual annotation makes it an unfeasible strategy for indexing a corpus the size of the parliamentary proceedings. 
On the other hand, while full-text search predominantly tracks specific expressions, it only captures fragments of the concepts we aim to find. In WideNet, querying (the net we throw out) should not just catch the big fish, even less should it rob the whole lake of its inhabitants; it should, as we argue below, obtain an \textit{optimal} and \textit{diverse} sample in a \textit{defensible} way.
Note that this formulation assumes the findings to have a certain margin of error, indicating the net might have some deficiencies: some of the fish might escape through its mazes. Even though we might miss some results, using imperfect instruments is still more convenient than hunting down each fish separately (i.e. using multiple queries).
To attenuate the distortions caused by technical errors (for example incorrectly linked entities), the user is given more agency to refine and adjust the scope and target of the query, making search an intrinsic part of the actual research. Even when the net is not a perfect instrument, by inspecting what it returns we can locate places which are fruitful for further exploration. 

\shrink
\subsubsection{It should be attractive to the intended users}
As stated previously, WideNet was designed to facilitate queries that comprise a wide range of entities, and thereby addresses a common problem encountered in historical research on vast digital corpora. Search engine evaluation traditionally assumes users to have specific information needs, which can be satisfied by proposing a ranking over documents. The information needs of historians, however, are often vaguely demarcated, and defining the scope of their interest is part of the research itself.

Humanist scholars generally pose a research question rather than a hypothesis. In this sense, an attractive interface understands the information need as a moving target.  For historians, search and research are inextricably intertwined practices: a constant feedback loop exists between selecting information and readjusting the query in the light of new evidence. 
The notions of ``source critique'' (procedures for evaluating documents) and ``heuristics'' (tools for finding and interpreting them) form the bedrock of historical research: Scrutinizing the past boils down to a continuous cycle of obtaining the relevant documents and interrogating them.
WideNet accommodates this search-as-research need by modeling the query as something to be sculpted during the research process. It allows users to judge the relevance of the found speeches at different levels of detail but also enables them to reverse any of the choices made.
Therefore, historians don't just have to dive into data but are guided by a search tool that functions as a dowsing rod that leads them to the most relevant spots in a gigantic dataset. 

%
%
Particularly for historians of the recent past, the amount of information easily becomes unmanageable for qualitative research. The rise of digitization and ``Big Data'' has merely reframed the problem (instead of solving it), especially for those historians who prefer close over distant reading---still by far the majority.
%
%
Queries often return more hits than one can realistically scrutinize. Given the flaws in current-day semantic technology, irrelevant results can constitute a large share of returned speeches and should only minimally distract the researcher.  
A traditional solution is to rank results by relevance. WideNet, however, explicitly entrusts users with more agency, by giving them a grouped overview of found speeches, so that even when confronted with thousands of hits, the scholar can separate the wheat from the chaff.

\shrink
\subsubsection{It should exploit the strengths of the machine and not compete with the strengths of the human}
Such a design choice hopes to align the strengths of algorithmic intelligence with expert judgment (instead of downgrading one at the expense of the other). This impacted both the selection and the presentation of material.

\textbf{Selection}: The classical trade-off between recall and precision also applies to academic search, but leads to different outcomes for humanist scholars. With respect to our task--exploring complex events--total recall remains unfeasible while prioritizing precision is undesirable. Unfeasible, because entity linkers are fallible; undesirable, because ultimately the responsibility for completeness and relevance of the found speeches is vested in the scholar and not the search algorithm, which requires us to be conservative when discarding elements as irrelevant, preferring expert judgment over algorithmic filtering.
%
%
%

\textbf{Presentation}: Given the limitations of EL systems and human attention, we designed the interface so that human and machine actually co-construct the final query in manageable steps. The filtering phase gives the user a ``vote'' to retain or discard general entities, which are clustered by category and accompanied by preview results. This dampens the gap in frequencies between the found entities, allowing users the possibility to grasp the variety of found elements. 
Here the user can narrow down the set of speeches to those who are most likely to relate to the research question at hand. Only later, after discarding the unwanted speeches, is the user encouraged to engage in close reading to establish whether the found texts contribute to solving the research question.
This procedure makes the ``long tail'' more visible, which ties in with the research practices of historians, for whom the anecdotal can have huge explanatory value. This is not to say that frequency is unimportant---it can signify the saliency of a concept---but it requires balancing since the rare stingy needles are usually of more interest than the evident blunt ones.

\shrink
\subsubsection{It should make sense economically}
Putting the researcher and machine in a position where they complement each other, enhances the cost-effectiveness of the search. WideNet significantly facilitates event exploration in two senses. Firstly, it shows what has \textit{not} been found by listing all relevant entities it queried but did not find, so the users won't lose time in pursuit of absent entities. The absence yields useful insights and satisfies the information need of the scholar in a negative sense.
Secondly, it replaces scrolling through pages of ranked lists by a stepwise iteration over grouped results. 
Thematic exploration dampens the imbalance between the highly and less frequent entities, making ``Big Data'' more manageable for qualitative historical research.  
It provides a more time and energy efficient way to assess large sets of search results because frequency and relevance can be inversely correlated, and because grouping the preview results gives the scholar an impression of the variety of the retrieved entities. Having such an overview adds context to the data exploration, and makes the search procedure somehow less linear, as the researcher has a sense of what is to come.
%
%
%

\shrink
\subsubsection{It should be clear to the users what the system can and cannot do}
The case studies have clearly shown that WideNet has its limitations: not all found entities and speeches turned out to be relevant. But instead of hiding these ``errors'' from users---cleaning the results to improve precision---we chose to make the deficiencies apparent and involve the user in the selection process. Thereby, scholars also learn what the system can and cannot do. Increasing transparency and allowing users to sculpt the search results, partly corrects (and justifies the use of) error-prone technologies for academic research: it makes researchers aware they should not take results at face value, and allows for further technological improvement. 

\shrink
\section{Opportunities for Improvement}
A fair amount of data is collected during the use of WideNet's interface, i.e., the initially selected target categories and time period, each decision about the relevance of categories and individual entities, as well as a relevance assessment of every document that is retrieved by the final query, are provided by the user and stored by the system. We are able to gather this data unobtrusively because the decisions that the WideNet user is asked to make are compatible with the corpus selection workflow. The data is expected to be valuable as gold standard for the (temporal) pruning of raw category trees, e.g. to enable automatic deselection of irrelevant entities, which would still be reversible by the user. Moreover, we aim for WideNet to learn to distinguish between contexts in which the same mentioned entity is either relevant or irrelevant for the user's ongoing research.

Speaking more generally, Entity Linking researchers and engineers benefit by engaging with a community of users who are intrinsically motivated and well able to closely inspect and assess the system output. Applications like WideNet enable us to inductively discover which entity links are useful for more user-oriented tasks. The current design does have the limitation that two kinds of errors are lumped together: entities may be deselected because of systematic EL errors, or because they are not relevant to the target of the search. Similarly, documents may be judged as irrelevant because the mentioned entity is not relevant in that particular context, or because it was incorrectly linked. We aim to learn to distinguish between these cases based on real-world usage data of WideNet, and will adapt the interface if needed.
Finally, to improve EL recall, we would need to extend the tool to allow documents to be added from manual queries. These `missed' results can help to point out the false negatives of the EL system as well as gaps in the KB coverage with respect to relevant entities in the document collection.

\shrink
\section{Conclusion}
In this paper we introduced WideNet, an entity-based interactive search tool, specifically tailored to assist historians with the exploration of large document collections. The main motivation behind the creation of WideNet was to show that \textit{(a)} even imperfect technologies such as the state-of-the-art Entity Linking systems can support useful niche applications, and \textit{(b)} tool design should take into account the methodological practices of the scholar.

With respect to the latter claim, we demonstrated how the interface supports historians by providing them with a holistic overview of references to complex phenomena such as historical events. Moreover, the grouped presentation of search results with different levels of context dampens frequency in favor of variety: long tail entities become more visible while highly frequent but irrelevant results can easily be discarded.
Lastly, WideNet allows the user to actively sculpture their queries along the way. The user can continuously revisit previously made choices via a bottom-up (qualitative) exploration of the corpus.

Our general conclusion is that the use case and application crucially determine whether the quality of an EL system is good enough or not: the same EL system may be useless for one task but very useful for another.  Hence, paraphrasing a famous paper on early machine translation \cite{Church1993}: there are good applications of crummy entity linkers.

\begin{acks}
We thank the anonymous reviewers for their insightful remarks.
This work was supported by the Netherlands Organization for Scientific Research (ExPoSe project, NWO CI \# 314.99.108).
\end{acks}

\bibliographystyle{ACM-Reference-Format}
\bibliography{sigproc} 

\end{document}